\documentclass[twocolumn,showpacs,preprintnumbers,amsmath,amssymb]{revtex4}
\usepackage{cases}
\usepackage{txfonts}
\usepackage{amssymb}
\usepackage{amsmath}

\usepackage{graphicx}

\usepackage{dcolumn}
\usepackage{bm}

\begin{document}
\title{Navigation in non-uniform density social networks}

\author{Yanqing Hu, Yong Li,  Zengru Di, Ying Fan\footnote{yfan@bnu.edu.cn}}
\affiliation{Department of Systems Science, School of Management and
Center for Complexity Research, Beijing Normal University, Beijing
100875, China}

\date{\today}

\begin{abstract}

Recent empirical investigations suggest a universal scaling law for the spatial structure of social networks. It is found that the probability density distribution of an individual to have a friend at distance $d$ scales as $P(d)\propto d^{-1}$. Since population density is non-uniform in real social networks, a scale invariant friendship network(SIFN) based on the above empirical law is introduced to capture this phenomenon. We prove the time complexity of navigation in 2-dimensional SIFN is at most $O(\log^4 n)$. In the real searching experiment, individuals often resort to extra information besides geography location. Thus, real-world searching process may be seen as a projection of navigation in a $k$-dimensional SIFN($k>2$). Therefore, we also discuss the relationship between high and low dimensional SIFN. Particularly, we prove a 2-dimensional SIFN is the projection of a 3-dimensional SIFN. As a matter of fact, this result can also be generated to any $k$-dimensional SIFN.

\end{abstract}

\maketitle

\section {Introduction}

To understand the structure of the social networks in which we live is a very interesting problem. As part of the recent surge of interest in networks, there have been active research about social networks\cite{Kleinberg-covergence,SWNature,
BA model,Human travel,Human travel commen,Mobile travel patten}. Besides some well known common properties such as small-world and community
structure\cite{full model navigation,Givens-Newman,Newman-review}, much attention has been dedicated to navigation in real social networks.

In the 1960s, Milgram and his co-workers conducted the first small-world experiment \cite{the oldest experiment}. Randomly chosen individuals in the United States were asked to send a letter to a particular recipient using only friends or acquaintances. The results of the experiment reveal that the average number of intermediate steps in a successful chain is about six. Since then, ``six degrees of separation" has became the subject of both experimental and theoretical research\cite{News,Play}. Recently,
Dodds \textit{et al} carried out an experiment study in a global social network consisting about 60,000 email users
\cite{recent experiment}. They estimated that social navigation can reach their targets in a median of five to seven steps, which is similar to the results of Milgram's experiment.

The first theoretical navigation model was proposed by Kleinberg\cite{navigation brief nature,navigation full}. He introduced an $n \times n $ lattice to model social networks. In addition to the links between nearest neighbors, each node $u$ is connected to a random node $v$ with a probability proportional to $d(u,v)^{-r}$, where $d(u,v)$ denotes the lattice distance between $u$ and $v$.
Kleinberg has proved that the optimal navigation can be obtained when the power-law exponent $r$ equals to $d$, where $d$ is the dimensionality
of the lattice, and the time complexity of navigation in that case is at most $O(\log^2 n)$. Since then, much attention has been dedicated to Kleinberg's navigation model\cite{Analyzing kleinberg,Oskar licentiate thesis,efficient routing}. Roberson \textit{et al.} studied the navigation problem in fractal networks, where they proved that $r=d$ was also the optimal power-law exponent in the fractal case\cite{Rob06}. Carmi, Cartozo and their cooperators have provided exact solutions respectively for the asymptotic behavior of Kleinberg's navigation model\cite{Asymptotic behavior,Extended Navigability}. More recently, the navigation probolem with a total cost restriction has also been discussed, where the cost denotes the length of the long-range connections\cite{Total cost,Total cost1}.

Meanwhile, recent empirical investigations suggest a universal spatial scaling law on social networks. Liben-Nowell \textit{et
al} explored the role of geography alone in routing messages within the LiveJournal social network\cite{Use Kleinberg search}. They found
that the probability density function (PDF) of geographic distance $d$ between friendship was about
$P(d)\propto d^{-1}$. Adamic and Ada also observed the $P(d)\propto d^{-1}$ law when investigating the Hewlett-Packard Labs email network\cite{power-law
networks Kleinberg search}.
Lambiotte \textit{et al} analyzed the statistical properties of a communication network constructed from the records of a mobile phone
company \cite{Renaud Lambiotte}. Their empirical results showed that the probability that two people $u$ and $v$ living at a geographic
distance $d(u,v)$ were connected by a link was proportional to $d(u,v)^{-2}$. Because the number of nodes having
distance $d$ to any given node is proportional to $d$ in 2-dimensional world, so the probability for an individual to have a friend at distance
$d$ should be $P(d)\propto d \cdot d^{-2}=d^{-1}$. More recently, Goldenberg \textit{et al} studied the effect of IT revolution on
social interactions\cite{distance}. Through analyzing an extensive data set of the Facebook online social network, they pointed out that social communication decrease inversely with the distance $d$ following the scaling law $P(d)\propto d^{-1}$ as well.

Such as in the LiveJournal social network, population density is non-uniform in real social networks\cite{Use Kleinberg search}. To deal with the navigation problem with non-uniform population density, a scale invariant friendship network (SIFN for short) model based on the above spatial scaling law $P(d)\propto d^{-1}$ of social networks is proposed in this paper. We prove the time complexity of navigation in a 2-dimensional SIFN is at most $O(\log^4 n)$, which indicates social networks is navigable. Dodds \textit{et
al} have pointed out that individuals often resort to extra information such as education and professional information besides geography location in the real searching experiment\cite{recent experiment}. Considering this phenomenon, navigation process in real world may be seen as the projection of navigation in a higher dimensional SIFN. Therefore, we further discuss the relationship between high and low dimensional SIFN. Particularly, we prove that a 2-dimensional SIFN can be seen as the projection of any $k$-dimensional SIFN($k>2$) through theoretical analysis.

\section {Navigation In Non-uniform Density Social Networks }

To deal with the non-uniform population density in real social networks, we divide the whole population into small areas and give the following two assumptions. First, the population density is uniform in each small area. Second, the minimum population density among the areas is $m$, while the maximum is $M$. We set $m>0$ to guarantee that a searching algorithm can always make some progress toward any target at every step of the chain.

Like Kleinberg's network (KN for short) and Liben-Nowell's
rank-based friendship network (RFN for short), we employ an $n \times
n $ lattice to construct SIFN. Without loss of generality, we assume each node $u$ has $q$ directed long-range connections, where $q$ is a constant\cite{navigation full}. To generate a long-range connection of node $u$, we first randomly choose a distance $d$ according to the observed scaling law $P(d)\propto d^{-1}$ in social networks. Then randomly choose a node $v$ from the node set, whose elements have the same lattice distance $d$ to node $u$, and create a directed long-range connection from $u$ to $v$. The lattice is assumed to be large enough that the long-range connections will not overlap.

For simplicity, we set $q=1$. Let $S$ denote the set of all nodes,
then the probability that $u$ chooses $v$ as its long-rang
connection in SIFN can be given by eq.(\ref{prob of SIF}).
\begin{equation}\label{prob of SIF}
Pr_\text{SIFN}(u,v)=\frac{1}{c(u,v)}\frac{d(u,v)^{-1}}{\sum_{d=1}^{n}d^{-1}}
\end{equation}
where $c(u,v)=|\{x|d(u,x)=d(u,v), x\in S\}|$ and $d(u,v)$ denotes the lattice distance between nodes $u$ and $v$. Likewise, the
probability that $u$ chooses $v$ as its long-rang connection in KN and RFN are given respectively by eq.(\ref{prob of K}) and
eq.(\ref{prob of RF}).
\begin{equation}\label{prob of K}
    Pr_\text{KN}(u,v,r)=\frac{d(u,v)^{-r}}{\sum_{w\neq u}d(u,w)^{-r}}
\end{equation}

\begin{equation}\label{prob of RF}
    Pr_\text{RFN}(u,v)=\frac{rank_u(v)^{-1}}{\sum_{w\neq u}rank_u(w)^{-1}}
\end{equation}
where $rank_u(v)=|\{w|d(u,w)<d(u,v), x\in S\}|$ denotes the
number of nodes within distance $d(u,v)$ to node $u$ in RFN\cite{navigation full,Use Kleinberg search}. Notice that, the number of nodes with a distance $d(u,v)$ in a $k$-dimensional($k>1$) lattice is proportional to $d{(u,v)^{k - 1}}$. Thus, a node $u$ connects to node $v$ with
probability proportional to $d(u,v)^{-a}$ does not mean
$P(d)\propto d^{-a}$ but $P(d)\propto d^{-a+k-1}$ instead. Therefore, ${Pr_\text{KN}(u,v,k)}$, ${Pr_\text{SIFN}(u,v)}$ and $Pr_\text{RFN}(u,v)$ are exactly the same for any $k$-dimensional lattice based network when population density is uniform.
However, SIFN always satisfies the empirical results $P(d)\propto d^{-1}$ in social networks compared with KN and RFN. Further, ${Pr_{KN}(u,v,k)}$,${Pr_\text{SIFN}(u,v)}$ and $Pr_\text{RFN}(u,v)$ can be quite different when the population density is non-uniform.

\begin{figure}
\center
\includegraphics[width=6cm]{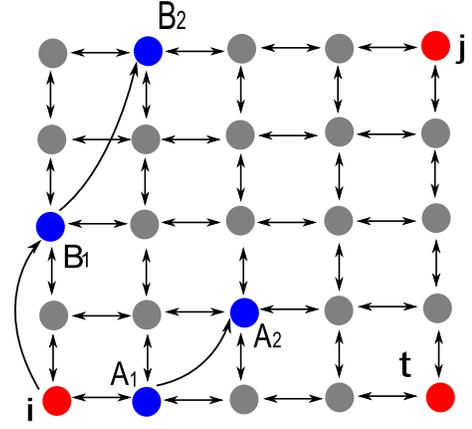}
\caption{Two strategies of sending message in a 2-dimensional SIFN.
Strategy $\mathcal{A}$, send the message directly to
target $t$ from the current message holder using
Kleinberg's greedy routing strategy. At each step, the message is sent to one of its neighbors who is most close to the target in the sense of lattice distance. Strategy $\mathcal{B}$, the message is first sent to a given node $j$ using Kleinberg's greedy strategy and then to the target node $t$ using the same strategy. Suppose we start from a source node $s$,
after one step, the message reaches nodes $A_1$ and $B_1$
respectively with strategy $\mathcal{A}$ and $\mathcal{B}$. Consider $B_1$ as the new
source node, then we should get $A_2$ and $B_2$ respectively with
strategies $\mathcal{A}$ and $\mathcal{B}$ in the next step. }\label{strategy}
\end{figure}

Since our 2-dimensional SIFN captures the non-uniform population density property in the real social networks, we purposefully divide the navigation process into two stages for simplicity. First send messages inside a small area and then among the areas. To analyze the time complexity of navigation in a 2-dimensional SIFN, we first compare the following two searching strategies as shown in FIG.\ref{strategy}. Strategy $\mathcal{A}$, send the message directly to target $t$ from the current message holder using Kleinberg's greedy routing strategy. At each step, the message is sent to one of its neighbors who is most close to the target in the sense of lattice distance. Strategy $\mathcal{B}$, the message is first sent to a given node $j$ using Kleinberg's greedy strategy and then to the target node $t$ using the same strategy. It can be proved that strategy $\mathcal{A}$ performs better than strategy $\mathcal{B}$ on average. Suppose we start sending message from the source node $s$, the message reaches nodes $A_1$ and $B_1$ respectively with strategy $\mathcal{A}$ and $\mathcal{B}$ after one step. It is always correct that lattice distance $d(A_1,t)$ $\leq$
$d(B_1,t)$, because greedy routing strategy always choose the node most close to target $t$ from its neighbors. According
to the results of\cite{Asymptotic behavior, efficient routing}, the longer the distance between a source and a given target, the more is the expected steps. Thus we should have $T(A_1\rightarrow t) \leq T(B_1\rightarrow t)$,
where $T(A_1\rightarrow t)$ and $T(B_1\rightarrow t)$ denote the
expected delivery time to target $t$ from $A_1$ and $B_1$ respectively.

Let $T(s\rightarrow j \rightarrow t)$ denote the expected
delivery time from $s$ to $t$ via a transport node $j$, then we have $T(s\rightarrow
t)\leq T(s\rightarrow B_1 \rightarrow t)$. Consider $B_1$ as a new
source node, then message will reach $A_2$ and $B_2$ with strategies $\mathcal{A}$ and $\mathcal{B}$ respectively in the next step. Following the same
deduction, we have $T(B_1\rightarrow A_2 \rightarrow t)\leq
T(B_1\rightarrow B_2 \rightarrow t)$. Repeat this process until the message
reaches the given node $j$ with strategy $\mathcal{B}$, then we should have a monotone increasing
sequence of expected delivery time \{$T(s\rightarrow B_1
\rightarrow t)$, $T(s\rightarrow B_2 \rightarrow t)$ , $\cdots$,
$T(s\rightarrow j \rightarrow t)$ \}. Therefore, we can obtain
$T(s\rightarrow t)$ $\leq$ $T(s\rightarrow j \rightarrow t)$, which
means strategy $\mathcal{A}$ is better than strategy $\mathcal{B}$. This
analysis can be extended to any $k$-dimensional SIFN.

Based on the first assumption and the fact that SIFN is identical to KN when population density is uniform, the expected steps spent in each small area using Kleinberg greedy algorithm is at most $O(\log^2 n)$. Consider each small area as a node, we will get a new 2-dimensional weighted lattice. The weight (population) of the nodes is between $m$ and $M$ based on the second assumption. Thus we have
\begin{equation}\label{bounds of probabilty}
c\frac{m}{M}d^{-1}\leq Pr_\text{SIFN}^{'}(u,v)\leq c\frac{M}{m}d^{-1}
\end{equation}
where $c$ is a constant and $Pr_\text{SIFN}^{'}(u,v)$ represents the probability that area $u$ is connected to area $v$ in the new weighted lattice.

We say that the execution of greedy algorithm is in
phase $j$ ($j>0$) when the lattice distance from the current node to target
$t$ is greater than $2^{j}$ and at most $2^{j+1}$.
Obviously, we have
\begin{equation}\label{4}
   \sum\limits_{d = 1}^n {d^{ - 1} }  \le 1 + \int\limits_1^n {x^{ - 1} dx = 1 + \log n < 2\log
   n}.
\end{equation}
Further, we define $B_j$ as the node set whose elements are within lattice distance $2^{j}+2^{j+1}<2^{j+2}$ to $u$.
Let $|B_j|$ denote the number of nodes in set $B_j$, we should have
\begin{equation}\label{5}
|B_j | > 1 + \sum\limits_{i = 1}^{2^j } {i > 2^{2j - 1} }.
\end{equation}
Suppose that the message holder is currently in phase $j$, then the probability that the node is connected by a long-range link to a node in phase $j-1$ is at least $(Mm^{ - 1} 2\log n \cdot 4 \cdot 2^{2j + 4} )^{ - 1}$.
The probability $\psi (x)$ to reach the next phase $j-1$ in more than $x$ steps can be given by
\begin{equation}\label{11}
\psi (x) = {(1 - {(M{m^{ - 1}}2\log n \cdot 4 \cdot {2^{2j + 4}})^{ - 1}})^x}
\end{equation}
and the average number of steps required to reach phase $j-1$ is
\begin{equation}\label{6}
< x >  = \sum\limits_{i = 1}^\infty  {{{(1 - \frac{m}{{256M\log n}})}^{i - 1}}}  = \frac{{256M\log n}}{m}.
\end{equation}
Since the initial value of $j$ is at most $\log n$, then the expected total number of steps required to reach the target is at most
$O(\frac{M}{m}\log^2 n)$.

As a matter of fact, it means that we are using strategy $\mathcal{B}$ to send message in 2-dimensional SIFN when the navigation process is divided into the above 2 stages. Thus, the time complexity of navigation in SIFN with strategy $\mathcal{B}$ is at most $O(\frac{M}{m}\log^4 n)$. However, actual navigation process in real world should be carried out regardless of the above two assumptions, which indicates individuals should use strategy $\mathcal{A}$. Based on the above analysis, strategy $\mathcal{A}$ performs better than strategy $\mathcal{B}$ on average. Therefore, the time complexity of navigation in 2-dimensional SIFN is at most $O(\log^4 n)$ with non-uniform population density.

\section{Relationship between high and low dimensional SIFN}

The empirical results show individuals always resort to extra information such as profession and education information besides the target's geography location when routing messages\cite{recent experiment}. Then, real navigation process in social networks may be modeled with a higher dimensional SIFN. In the following, we  will discuss the relationship between the high and low dimensional SIFN and prove that a 2-dimensional SIFN can be obtained by any $k$-dimensional SIFN ($k>2$). Particularly, we will provide the theoretic analysis for the case where $k=3$. The analysis can be generated to any $k$ dimensional cases.

We employe a random variable $D_3$ to denote the friendship distance in a 3-dimensional SIFN. For simplicity, a continuous expressions is used. Since, the long-range connections in 3-dimensional SIFN satisfies the above empirical law, the PDF of $D_3$ can be expressed by
\begin{equation}\label{PDF of D_3 }
    P(D_3=d)=\frac{1}{\ln d_{M} -\ln d_{m}}\frac{1}{d}, d_{m}\leq d\leq d_{M}
\end{equation} where $d_{m}$ and $d_{M}$ denote the minimum and maximum distance respectively in the 3-dimensional SIFN.

We can obtain a 2-dimensional network model if we project a 3-dimensional SIFN to a 2-dimensional world. Similarly, a random variable $D_2$ is used to denote the friendship distance in the new 2-dimensional network model. It is not difficult to understand that the condition for a 2-dimensional SIFN should be the PDF of $D_2$ satisfies $P(d)\propto d^{-1}$. Since $D_2$ is the projection of $D_3$, then $D_2$ can be seen as the product of $D_3$ and $X$. Here random variable $X$ is independent on $D_3$ and its PDF can be given by eq.(\ref{PDF of x}).

\begin{equation}\label{PDF of x}
    P(X=x)=\frac{1}{\lambda} , 0\leq x\leq \lambda
\end{equation}
where $0\leq \lambda \leq1$. Finally, the PDF of $D_2$ can be written as

\begin{equation}\label{PDF of D_2}
P(D_2=d) =\begin{cases}
0,  &\text{$d \leq 0$} \\
\frac{d_{M}-d_{m}}{d_{M}d_{m}\lambda(\ln d_{M} -\ln d_{m})}, &\text{$0<d \leq d_{m} \lambda$}\\
\frac{\frac{1}{d} - \frac{1}{d_{M}\lambda}}{\ln d_{M} -\ln d_{m}}, &\text{$d_{m} \lambda <d \leq d_{M} \lambda$}\\
0,  &\text{$d>d_{M} \lambda$}
\end{cases}
\end{equation}
When taking account of real social networks, $d_{M}$ is large
enough that the term $\frac{1}{d_{M} \lambda}$  will approach its limit of 0. Meanwhile, the term $d_{m}\lambda$ can be
neglected when compared with $d_{M}\lambda$, because $\lambda\leq 1$ and
$d_{m}$ is relatively small. Thus the PDF of $D_2$
can be simplified into $P(d)\propto d^{-1}$, which
is identical to that of $D_3$ in a 3-dimensional SIFN.

Through theoretical analysis, we have proved a 2-dimensional SIFN can be seen as the projection of a 3-dimensional SIFN. Likewise, we can get a 2-dimensional SIFN from any $k$-dimensional($k>2$) SIFN. Notice that individuals are always restricted on the 2-dimensional geography world even they possess extra information from other dimensions. Thus, real-world searching process may be seen as the projection of navigation in a high dimensional SIFN. Our analysis indicate that SIFN model may explain the navigability of real social networks even take account of the fact that individuals always resort to extra information in real searching experiments.

\section {Conclusion}

Recent investigations suggest that the probability distribution of having a friend at distance $d$ scales
as $P(d)\propto d^{-1}$. We propose an SIFN model based on this spatial property to deal with navigation problem with non-uniform population density. It has been proved that the time complexity of navigation in 2-dimensional SIFN is at most $O(\log^4 n)$, which corresponds to the upper bond of navigation in real social networks. Given the fact that individuals are always restricted on the 2-dimensional geography world even they possess information of the higher dimensions, actual searching process can be seen as a projection of navigation in a higher $k$-dimensional SIFN. Through theoretical analysis, we prove that the projection of a higher $k$-dimensional SIFN results in a 2-dimensional SIFN. Therefore, SIFN model may explain the navigability of real social networks even take account of the information from higher dimensions other than geography dimensions.

\section {Acknowledgement}

We thank Prof. Shlomo Havlin for
some useful discussions. This work is partially supported by the Fundamental Research Funds for the Central Universities and NSFC under
Grants No.70771011 and No. 60974084 and NCET-09-0228. Yanqing Hu is supported by Scientific Research Foundation and Excellent Ph.D Project of Beijing Normal University.

\end{document}